\pdfoutput=1

\documentclass[11pt]{article}

\usepackage[final]{acl}

\usepackage{times}
\usepackage{latexsym}

\usepackage[T1]{fontenc}

\usepackage[utf8]{inputenc}

\usepackage{microtype}

\usepackage{inconsolata}

\usepackage{graphicx}

\usepackage{algpseudocode}
\usepackage[english]{babel}
\usepackage[utf8]{inputenc}
\usepackage{tcolorbox}
\usepackage{enumitem}
\usepackage{pifont}
\usepackage[normalem]{ulem}
\usepackage{booktabs}
\usepackage{bm}
\usepackage{subcaption}
\usepackage{adjustbox}
\usepackage{multirow}
\usepackage{amsmath}
\usepackage{amsmath,amssymb,amsfonts}
\usepackage{algorithm}

\usepackage{setspace}
\usepackage{textcomp}
\usepackage{xcolor,xspace}
\newcommand{\ours}{TrickCatcher\xspace}

\def\ie{{i.e.,} }

\usepackage[framemethod=tikz]{mdframed}
\newcounter{finding}
\newcommand{\finding}[1]{\refstepcounter{finding}
  \vspace{1.5mm}
 \begin{mdframed}[linecolor=gray,roundcorner=12pt,backgroundcolor=gray!15,linewidth=3pt,innerleftmargin=2pt, leftmargin=0cm,rightmargin=0cm,topline=false,bottomline=false,rightline = false]
 
  \textbf{Ans. to RQ\arabic{finding}:} #1
 \end{mdframed}
 \vspace{1.5mm}
}

\title{LLM-Powered Test Case Generation for Detecting Bugs in \\Plausible Programs}

\author{
 \textbf{Kaibo Liu\textsuperscript{1}},
 \textbf{Zhenpeng Chen\textsuperscript{2,}\thanks{Corresponding authors}},
 \textbf{Yiyang Liu\textsuperscript{1}},
 \textbf{Jie M. Zhang\textsuperscript{3}},
 \textbf{Mark Harman\textsuperscript{4}},\\
 \textbf{Yudong Han\textsuperscript{1}},
 \textbf{Yun Ma\textsuperscript{1}},
 \textbf{Yihong Dong\textsuperscript{1}},
 \textbf{Ge Li\textsuperscript{1,}\footnotemark[1]},
 \textbf{Gang Huang\textsuperscript{1,5}}
\\
\setstretch{0.5}
 \textsuperscript{1}Peking University,
 \textsuperscript{2}Nanyang Technological University,\\
 \textsuperscript{3}King's College London,
 \textsuperscript{4}University College London\\
 \textsuperscript{5}National Key Laboratory of Data Space Technology and System
\\
\setstretch{0.5}
\texttt{\{liukb,hanyd,mayun,lige,hg\}@pku.edu.cn,zhenpeng.chen@ntu.edu.sg}\\
\texttt{ \{ptr1479,dongyh\}@stu.pku.edu.cn,jie.zhang@kcl.ac.uk,mark.harman@ucl.ac.uk}
}

\begin{document}
\maketitle
\begin{abstract}
Detecting tricky bugs in plausible programs, those that pass existing test suites yet still contain bugs, remains a significant challenge in software testing. To address this problem, we propose \ours, an LLM-powered approach to generating test cases for uncovering bugs in plausible programs. \ours operates in three stages: First, it uses an LLM to generate program variants based on the program under test (PUT) and its specification. Second, it employs an LLM to construct an input generator from the specification for producing test inputs. Finally, these inputs are executed on both the PUT and its program variants to detect inconsistencies in their outputs. We evaluate \ours on two datasets, TrickyBugs and EvalPlus, which include 366 human-written and 151 AI-generated plausible programs with tricky bugs. \ours achieves recall, precision, and F1 scores that are 1.80×, 2.65×, and 1.66× those of the state-of-the-art baselines, respectively. Code and data used are available at \url{https://github.com/RinCloud/TrickCatcher}.
\end{abstract}

\section{Introduction}
\label{sec:intro}
Validating that programs meet a given specification that defines their intended functionality is critical, with software testing serving as the primary approach to achieving this goal. Central to software testing lies in the use of a \emph{test suite}, a collection of test cases designed to validate the program under test (PUT). Each test case consists of a \textit{test input}, sampled from the input space of the PUT, and a \textit{test oracle}, which specifies the expected output based on the program’s specification.

\begin{figure}[t]
  \centering
  \includegraphics[width=1\columnwidth]{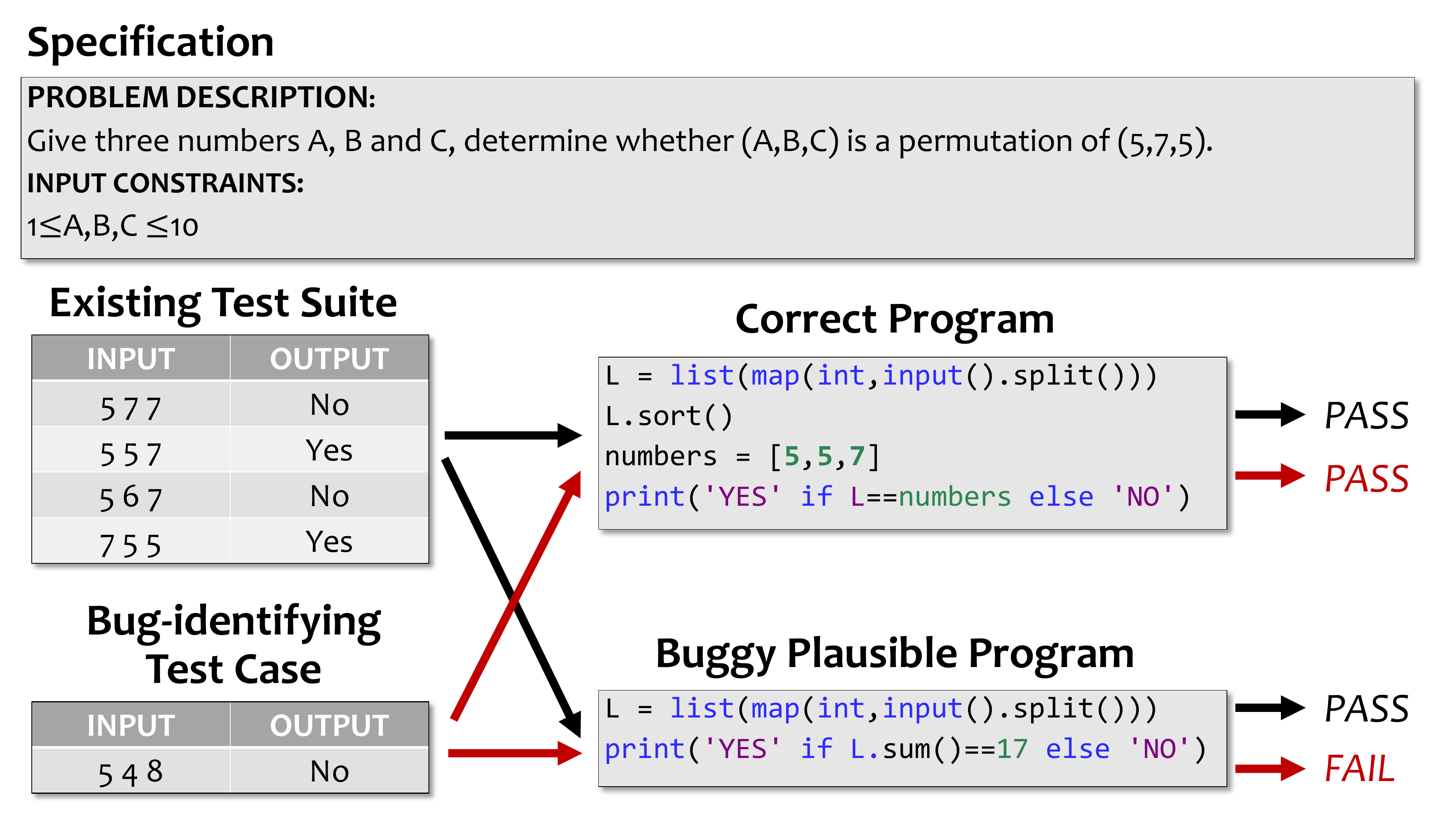}
  \caption{A motivating example.}
  \label{fig:motivating_example}
\end{figure}

Programs that pass all test cases are considered \textit{plausible programs}~\cite{liu2024trickybugs,B4}, but plausibility does not equate to correctness. Such plausible programs may still harbor subtle bugs~\cite{liu2023oj,silent2024,counterfeit}, often logical corner cases, that escape detection by test suites. We refer to these elusive bugs as \textit{tricky bugs}. For example, Figure~\ref{fig:motivating_example} shows a real-world plausible program from an online judge platform\footnote{\url{https://atcoder.jp/contests/abc042/tasks/abc042_a}} that passes an existing test suite yet conceals a tricky bug. 
A correct program should check whether the set of the three input numbers is \{5, 7, 5\}, while the buggy plausible program only checks if the sum of the three numbers is 17.

In fact, tricky bugs are surprisingly common. A recent study~\cite{liu2023oj} identified 3,440 such bugs in human-written programs deemed correct by Online Judge platforms, underscoring their prevalence even in scenarios where code has been thoroughly reviewed and tested. Despite their significant impact, existing research has largely overlooked the development of testing approaches specifically designed to uncover tricky bugs in plausible programs, leaving a critical gap in current testing practices.

To fill the gap, we propose \ours, an LLM-powered test case generation approach for detecting tricky bugs in plausible programs. \ours combines LLMs and differential testing to accurately generate test inputs and test oracles. 
It consists of three steps: \textit{program variant generation}, \textit{test input generation}, and \textit{differential testing}. In the first two steps, we use LLMs to generate various program variants and test inputs of the PUT. In the third step, we continually feed generated test inputs to both the PUT and program variants, searching for inconsistencies in program outputs. 

A straightforward approach is to directly use the specification as input for LLMs to generate program variants and test inputs. However, our preliminary experiments reveal two major limitations of this naive approach: 
\ding{182} \emph{Low correctness of program variants.} While LLMs can generate correct programs for simple tasks based on specifications~\cite{humaneval}, their performance deteriorates with complex tasks (e.g., competition-level programs). The resulting variants often contain errors, reducing the effectiveness of differential testing. 
\ding{183} \emph{Low correctness of test inputs.} Although LLMs can generate valid test inputs for simple formats (e.g., two integers), they struggle with inputs requiring constraints (e.g., a square matrix with monotonically increasing rows). Our experiments show that when test inputs are generated directly from specified input constraints, 40.10\% of the generated test inputs are invalid. Invalid inputs fail to expose bugs, as they result in undefined program behavior, and worse, they can cause \textit{false positives}, incorrectly marking a working program as buggy.

To tackle these challenges, we propose the following three solutions in \ours: 

\noindent \ding{182} \textbf{PUT-guided program variant generation.} Instead of relying solely on specifications, \ours provides the PUT alongside the specification in the LLM prompt. The LLM is tasked with analyzing the PUT and generating repaired program variants if necessary. These variants are then filtered using existing test cases to exclude those that fail. By leveraging the PUT as a foundation, \ours steers the LLM toward generating meaningful modifications rather than creating implementations from scratch, significantly improving the quality of the program variants.

\noindent \ding{183} \textbf{Generator-based input generation.} Instead of directly generating test inputs from specifications, \ours instructs the LLM to create an input generator (e.g., a Python script) that adheres to specified constraints and then uses the generator to generate test inputs. This approach separates logical reasoning from input generation, allowing \ours to overcome the reasoning limitations of LLMs. As a result, it significantly improves the validity of the generated inputs, achieving a higher proportion of valid test cases.

\noindent \ding{184} \textbf{Diversity-driven differential testing.} Recognizing that program variants can inherit similar bugs from the PUT, \ours departs from the traditional majority-voting principle, which assumes the most frequent output to be correct. Instead, it prioritizes diversity in test outputs to construct test oracles. This counterintuitive approach enhances the ability to detect subtle discrepancies and uncover tricky bugs that majority voting might overlook.

We evaluate \ours on two datasets, TrickyBugs~\cite{liu2024trickybugs} and EvalPlus~\cite{evalplus}, which contain 366 human-written and 151 AI-generated plausible programs with tricky bugs, respectively.  TrickyBugs includes both C++ and Python programs, while EvalPlus focuses on Python. 
\ours is compared against three representative baselines, and the results demonstrate its superior performance in recall, precision, and F1 score, achieving up to \bm{$1.80\times$}, \bm{$2.65\times$}, and \bm{$1.66\times$} of the best baseline, respectively.
In particular, \ours achieves F1 scores of 41.31\%, 42.35\%, and 51.34\% on TrickyBugs (C++), TrickyBugs (Python), and EvalPlus, significantly outperforming the best baseline's F1 scores of 24.95\%, 36.20\%, and 35.76\%. 
An ablation study further confirms that each component of \ours contributes meaningfully to its overall performance.

\section{Related Work}
\label{sec:relatedWork}

\noindent \textbf{Traditional test case generation.} Traditional test case generation methods primarily rely on search-based~\cite{survey_sbst,survey_sbst2} and symbolic execution-based approaches~\cite{survey_symbolic_execution}, with popular tools such as EvoSuite~\cite{evosuite}, Pynguin~\cite{pynguin}, and KLEE~\cite{klee} exemplifying these approaches. However, these traditional approaches cannot automatically parse program specifications, which are often written in natural language and crucial for bug detection. In contrast, \ours leverages LLMs to interpret and utilize program specifications effectively.

\noindent \textbf{LLM-based test case generation.}
Recently, LLMs have been widely adopted in test case generation approaches~\cite{DBLPabs240902977}, such as ChatTester~\cite{NoMoreManualTests}, TestPilot~\cite{Testpilot}, ChatUnitTest~\cite{chatunitest}, and SymPrompt~\cite{symprompt}. However, these approaches differ from \ours as they primarily focus on improving test coverage rather than detecting bugs. 
In terms of bug detection, Differential Prompting (DP)~\cite{nuance} represents the state-of-the-art in LLM-based test case generation. As a differential testing approach, DP also generates program variants to identify potential bugs. The key differences between DP and \ours are as follows: (1) \ours focuses on plausible programs, which enables it to consider the existing test suite when generating inputs, whereas DP, which is not designed for plausible programs, does not; (2) \ours uses both the PUT and its specification for program variant generation, while DP relies solely on the inferred specification;  (3) \ours employs LLMs to generate input generators, which are then used to produce test inputs, in contrast to DP, which directly generates inputs from the specification; and (4) \ours introduces diversity-driven differential testing, whereas DP uses majority voting, a traditional approach in differential testing, for test oracle construction. We have implemented a variant of DP to make it applicable to plausible programs and compared it with \ours. Moreover, we thoroughly evaluate these designs of \ours in Section \ref{sec:ablation}.

\section{Problem Definition}
\label{sec:prelimi}
In this section, we define the problem of generating test cases to detect bugs in plausible programs.

Given a program specification \(S\), it defines an intended mapping \(f\) from the input space \(I\) to the output space \(O\). For any input \(in \in I\), the correct output (test oracle) is \(out = f(in)\). The goal of bug-identifying test case generation is to generate an input \(in_t\) along with its corresponding correct output \(f(in_t)\), forming the test case \((in_t, f(in_t))\), such that the program under test (PUT) produces an incorrect output, \ie \(f_{PUT}(in_t) \neq f(in_t)\). 

If a program \(P_0\) passes all test cases in a test suite \(T_0\), we call \(P_0\) a \textit{plausible program} (relative to \(T_0\)). However, if \(P_0\) still contains a bug, we refer to it as a \textit{buggy plausible program}~\cite{B4, counterfeit}, and the bugs in \(P_0\) are termed \textit{tricky bugs}~\cite{liu2023oj,liu2024trickybugs}.

The objective of this paper is to generate test cases for a plausible program \(P_0\) based on its specification \(S\). A failed test case is considered a potential bug identifier as it reveals a discrepancy between the program's output and the expected output (\ie \(f_{PUT}(in_i) \neq f(in_i)\)). However, not every failed test case necessarily indicates a bug in the program, as failures can also result from errors in the test case itself.

For a test case \((in, out)\) to be valid, it must meet the following two conditions: \textbf{(1) Validity of the test input:} The input \(in\) must belong to the valid input space \(I\). \textbf{(2) Correctness of the test oracle:} The output \(out\) must satisfy \(out = f(in)\), where \(f\) is the mapping specified by \(S\).

If a test case \((in, out)\) fails, and both conditions are satisfied, the test case is a \textit{True Positive} (TP), indicating the presence of a bug in the PUT. If the test case fails to meet one or both conditions, it results in a false alarm, which is categorized as a \textit{False Positive} (FP).

\section{Our Approach: \ours}
\label{sec:methodology}

\begin{figure*}[!htbp]
  \centering
  \includegraphics[width=1\linewidth]{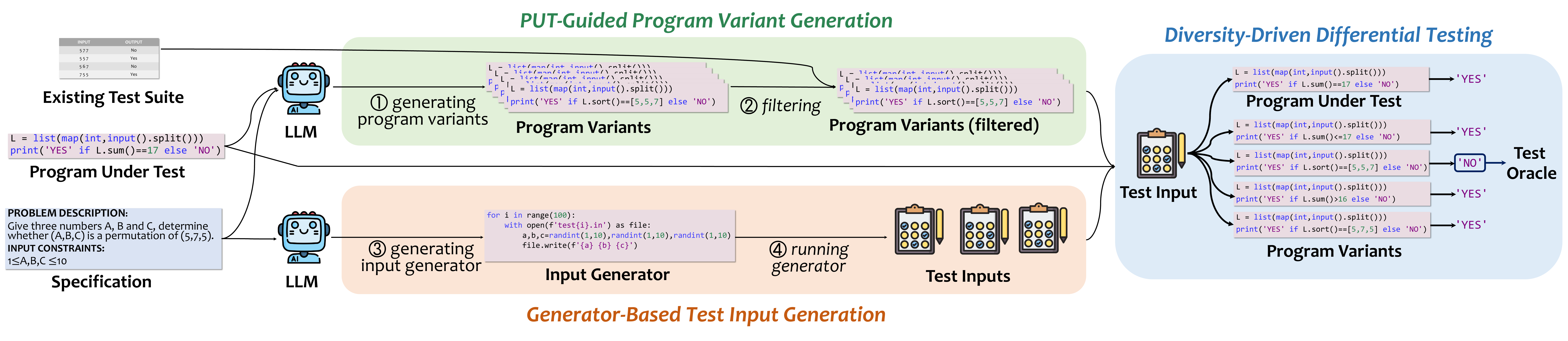}
  \caption{Overview of \ours.}
  \label{fig:overview_of_AID}
\end{figure*}

Figure \ref{fig:overview_of_AID} provides an overview of \ours. It takes the program specification, the PUT, and the existing test suite as inputs, and outputs a set of test cases designed to detect bugs in the PUT. The workflow of \ours consists of three key steps: PUT-guided program variant generation, generator-based test input generation, and diversity-driven differential testing. The first two steps use LLMs to generate multiple program variants and test inputs for the PUT, while the third step iteratively feeds these test inputs into both the PUT and its program variants, searching for inconsistencies in the outputs and constructing potential bug-identifying test cases. Each step is explained in detail below.

\subsection{Program Variant Generation}
\label{sec:put-guided_progen}
\begin{figure}[t]
  \centering
  \includegraphics[width=1\linewidth]{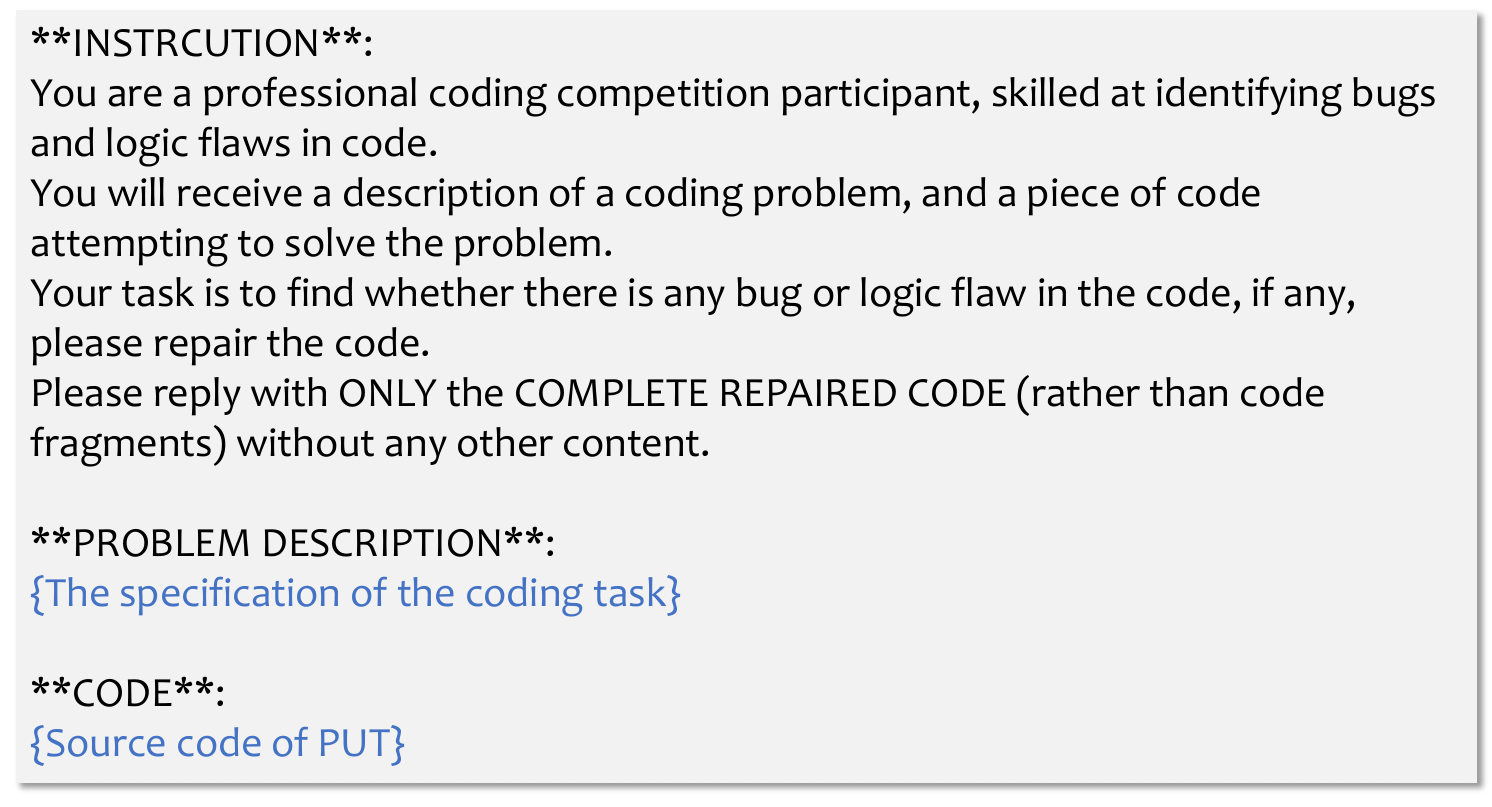}
  \caption{Prompt for generating program variants.}
  \label{fig:AID_prompt_program_generation}
\end{figure}

The first step of \ours involves generating program variants of the PUT. Both the program specification and the PUT are provided to the LLM, which is prompted to assess whether the PUT contains any bugs based on the specification. If the LLM detects a potential bug, it is tasked with repairing the program and generating a corrected version. The prompt used is illustrated in Figure \ref{fig:AID_prompt_program_generation}.

The key advantage of PUT-guided program generation is its ability to leverage both the program specification and the PUT. Since the PUT is a plausible program, it already exhibits a certain level of correctness, particularly for the input space covered by the existing test suite. By generating variants based on the PUT, the LLM is more likely to produce high-quality variants with a reduced risk of introducing new bugs compared to generating variants directly from the specification alone.

To improve the correctness of the program variants, \ours filters out any variants that fail the existing test suite. This filtering step makes effective use of the information provided by the test suite, ensuring that only high-quality program variants are retained for the subsequent steps of differential testing.

\subsection{Test Input Generation}
\label{sec:generator-based}
\begin{figure}[htbp]
  \centering
  \includegraphics[width=1\linewidth]{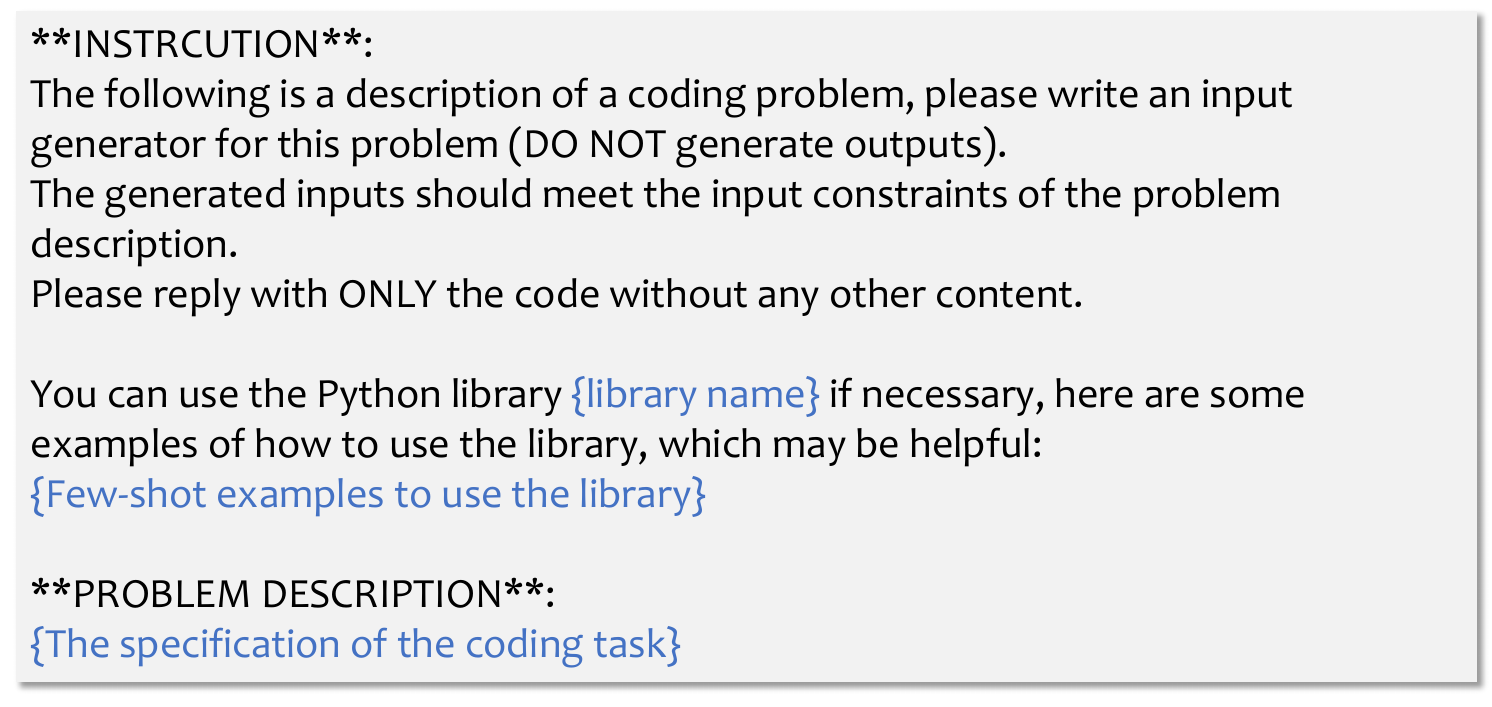}
  \caption{Prompt for generating test input generator.}
  \label{fig:AID_prompt_test_generator_generation}
\end{figure}

The second step of \ours involves generating test inputs. The main challenge here is ensuring that the generated test inputs are valid, meaning that they satisfy the required input constraints.

Directly generating test inputs based on the specification using LLMs often results in invalid inputs, as LLMs have limited reasoning capabilities, particularly when the specified constraints are complex. To address this challenge, we adopt a two-step approach: first, we prompt the LLM to summarize the constraints and translate them into code. Then, we use the generated code to produce valid test inputs.

Specifically, \ours proposes generator-based input generation. The LLM is tasked with creating a test input generator, which is then executed to produce the inputs. In our approach, the generator is specified as a Python script, as shown in the prompt in Figure~\ref{fig:AID_prompt_test_generator_generation}. To enhance the generator's capabilities, we can provide the LLM with a library of functions through few-shot learning examples, enabling it to efficiently learn how to use the library. We choose the Python library \verb|CYaRon| in our experiment, and the library can be easily replaced by adjusting the few-shot examples.

\begin{algorithm}[htb]\label{alg:div}
\caption{Diversity-driven differential testing}\label{alg:diff_testing}
\textbf{INPUT:}   
PUT, Set of program variants $\mathcal{P}$, Set of generated inputs $\mathcal{I}$\\
\textbf{OUTPUT:} 
Test cases

\begin{algorithmic}[1]
\small
\State{$\mathit{testCases} \gets \emptyset$}
\For {$\mathit{input} \in\mathcal{I}$}
  \State{$\mathit{diff} \gets \emptyset$}
  \State $\mathit{output}_0 \gets f_{PUT}(\mathit{input})$
  \For {$\mathit{P} \in \mathcal{P}$}
    \State {$\mathit{output} \gets f_P(\mathit{input})$}
    \If {$\mathit{output} \neq \mathit{output}_0$} 
      \State {Add $\mathit{output}$ to $\mathit{diff}$}
    \EndIf
    \EndFor
    \If {$\mathit{diff} \neq \emptyset$}
      \State {$\mathit{oracle} \gets$ most frequent element in $\mathit{diff}$}
      \State {$\mathit{test} \gets (\mathit{input},\mathit{oracle})$}
      \State {Add $\mathit{test}$ to $\mathit{testCases}$}
    \EndIf
\EndFor
\State {\textbf{return} $testCases$}
\end{algorithmic}
\end{algorithm}

\subsection{Differential Testing}
\label{sec:diversity-first}
The third step of \ours involves constructing a test oracle through differential testing. As shown in Figure \ref{fig:overview_of_AID}, \ours feeds the generated test inputs to both the PUT and the generated program variants, searching for inconsistencies in their outputs. 

The algorithm is detailed in Algorithm~\ref{alg:diff_testing}. \ours introduces diversity-driven differential testing: if a program variant produces an output different from the PUT's output, we take the variant's output as the test oracle (the correct output). If multiple outputs differ from the PUT's, the most frequent output is selected as the oracle. If all outputs match the PUT, the input is discarded, and the next input is tested.

This approach is counterintuitive, as developers typically rely on majority voting in differential testing~\cite{liu2023oj}. The rationale behind our algorithm is that while the LLM may correctly replicate parts of the PUT, it can also be misled by the PUT. As a result, program variants may inherit the same bugs as the PUT. Our experiments show that it is common for some variants to produce the same erroneous output as the PUT. Thus, we place greater trust in program variants that differ from the PUT’s output.

\section{Evaluation}
\label{sec:evaluation}

\subsection{Research Questions (RQs)}
We aim to comprehensively evaluate \ours by answering the following RQs.

\noindent \textbf{RQ1:} How effective are the test cases generated by \ours in detecting bugs in plausible programs?

\noindent \textbf{RQ2:} How many false positives does \ours generate when applied to correct programs?

\noindent \textbf{RQ3:} How do the different components contribute to the final performance of \ours?

\noindent \textbf{RQ4:} How does the number of program variants impact the effectiveness of \ours?

\noindent \textbf{RQ5:} How does the difficulty of coding tasks impact the effectiveness of \ours?

\begin{table*}[t]
\centering
\caption{(RQ1) Effectiveness of different methods in detecting bugs in plausible programs. $k$ denotes the number of generated program variants. R, P, and F denote recall, precision, and F1 score. Bold numbers indicate the best F1 scores. Overall, \ours shows the best effectiveness in detecting bugs in both human-written and AI-generated plausible programs.}
\label{tab:effectiveness_all}
\begin{adjustbox}{width=\textwidth}
\begin{tabular}{llrrr@{\hspace{10mm}}rrr@{\hspace{10mm}}rrr}
\toprule
\multirow{2}{*}{} & \multirow{2}{*}{$k$} & \multicolumn{3}{c}{TrickyBugs (C++)} & \multicolumn{3}{c}{TrickyBugs (Python)} & \multicolumn{3}{c}{EvalPlus}\\ \cmidrule(l){3-5} \cmidrule(l){6-8}\cmidrule(l){9-11}
 &  & R & P & F & R & P & F& R & P & F\\ \midrule
 
\texttt{CHAT}  & -   & 3.78  & 6.31  & 4.27  & 3.77  & 8.85  & 5.29  & 1.21 & 8.28 & 2.12 \\ \midrule
\texttt{APR}  & -    & 16.46 & 34.58 & 22.30 & 10.20 & 36.64 & 15.96 & 41.39 & 49.97 & 45.28 \\ \midrule

\multirow{5}{*}{\texttt{DPP}} & 2  & 20.96 & 26.37 & 23.35 &  32.54 & 40.79 & 36.20 & 23.36 & 53.54 & 32.52 \\
                              & 4  & 19.46 & 32.22 & 24.27 &  28.72 & 46.32 & 35.46 & 23.01 & 61.16 & 33.45 \\
                              & 6  & 17.68 & 41.60 & 24.81 &  26.17 & 53.03 & 35.05 & 22.80 & 72.71 & 34.71\\
                              & 8  & 16.31 & 53.09 & 24.95 &  24.41 & 62.02 & 35.03 & 22.55 & 80.51 & 35.23\\ 
                              & 10 & 15.50 & 60.80 & 24.71 &  23.22 & 70.26 & 34.90 & 22.29 & 90.36 & 35.76\\ \midrule
\multirow{5}{*}{\texttt{\ours}} & 2   & 25.06 & 69.91 & 36.90 & 23.74 & 78.95  & 36.51 & 32.78 & 85.09 & 47.33 \\
                                     & 4   & 27.74 & 69.69 & 39.68 & 27.17 & 77.81  & 40.28 & 35.35 & 84.41 & 49.83 \\
                                     & 6   & 28.65 & 69.45 & 40.57 & 28.69 & 77.81  & 41.92 & 36.32 & 83.85 & 50.69 \\
                                     & 8   & 29.19 & 69.34 & 41.09 & 29.09 & 77.81  & 42.35 & 36.90 & 83.38 & 51.16\\ 
                                     & 10  & 29.38 & 69.57 & \textbf{41.31} & 29.09 & 77.81 & \textbf{42.35} & 37.14 & 83.14 & \textbf{51.34}\\ \midrule
\multirow{3}{*}{Improvement}      
&  Average &  55.73\% $\uparrow$ &  62.54\% $\uparrow$  &  63.44\% $\uparrow$  &  2.01\% $\uparrow$  &  43.23\% $\uparrow$  &  15.16\% $\uparrow$  &  56.56\% $\uparrow$ &  17.19\% $\uparrow$  &45.83\% $\uparrow$ \\
&  Best vs. Best  & 80.13\% $\uparrow$ & 31.04\% $\uparrow$ & 65.57\% $\uparrow$ & 10.61\% $\downarrow$ & 90.76\% $\uparrow$ & 16.99\% $\uparrow$ & 66.62\% $\uparrow$ & 7.99\% $\downarrow$ & 43.57\% $\uparrow$ \\
&  Worst vs. Worst & 19.56\% $\uparrow$ & 165.11\% $\uparrow$ & 58.03\% $\uparrow$ & 2.24\% $\uparrow$ & 12.37\% $\uparrow$ & 4.61\% $\uparrow$ & 40.33\% $\uparrow$ & 58.93\% $\uparrow$ & 45.50\% $\uparrow$ \\

\bottomrule
\end{tabular}
\end{adjustbox}
\end{table*}

\subsection{Datasets}
\label{sec:dataset}
We evaluate \ours using two datasets: \emph{TrickyBugs}~\cite{liu2024trickybugs}, which tests its ability to detect bugs in human-written plausible programs, and \emph{EvalPlus}~\cite{evalplus}, which assesses its effectiveness on AI-generated plausible programs.\footnote{The two datasets are under the MIT and Apache 2.0 licenses, respectively.} In total, we use 366 human-written and 151 AI-generated plausible programs as PUTs. Below are brief descriptions of the two datasets:

\noindent $\bullet$ \textbf{TrickyBugs} contains hundreds of coding tasks from an online judge platform, with plausible programs submitted by real participants. Although these programs pass the existing test suite, they contain bugs, and the dataset provides additional test cases to detect them. We use 251 C++ and 115 Python plausible programs from this dataset.

\noindent $\bullet$  \textbf{EvalPlus} is a code generation benchmark including 164 Python coding tasks, each with base and extra test cases. We filter programs from EvalPlus’s pre-generated LLM code samples\footnote{\url{https://github.com/evalplus/evalplus/releases/tag/v0.1.0}} to obtain buggy plausible programs that pass the base test cases but fail the extra ones. The final set includes 151 coding tasks, each with an AI-generated plausible program.

\subsection {Evaluation Metrics}
\label{sec:mertrics}
We have defined TP and FP in Section \ref{sec:prelimi}. To distinguish between TPs and FPs, we need to know the correct outputs and validity of the generated test inputs. For correct outputs, both datasets provide canonical programs, and we use the outputs from these canonical programs as the reference for correctness. To validate the test inputs, we use the provided Python checkers for the EvalPlus dataset, while for the TrickyBugs dataset, we manually verify the input validity.

We further define precision, recall, and F1-score:

\noindent $\bullet$ \textbf{Precision} is defined as $\frac{\# \textit{TP}}{\# \textit{TP} + \# \textit{FP}}$. It determines the practicality of test generation approaches~\cite{towardsMoreRealistic}. The higher the precision, the fewer false positive test cases developers need to check before confirming a true bug.

\noindent $\bullet$ \textbf{Recall} is defined as $\frac{\# \textit{TP}}{\# \textit{TP} + \# \textit{FN}}$. Note that for buggy PUTs, all negatives are false negatives; for correct PUTs, all negatives are true negatives.

\noindent $\bullet$ \textbf{F1 score} is defined as $\frac{2\times\textit{Precision}\times\textit{Recall}}{ \textit{Precision} + \textit{Recall}}$. Since recall and precision often exhibit an inverse relationship, the F1-score provides a harmonic mean to balance them, making it a widely used metric.

To ensure the reliability of our results, we conduct multiple runs of our experiments and calculate the average metric values across these runs. Detailed information on the repetition methods is provided in the Appendix~\ref{sec:repetition}.

\subsection{Baselines}
\label{sec:baseline}
We use three representative methods as baselines to compare with \ours (\texttt{TC}).

\noindent $\bullet$ \textbf{DirectChat (\texttt{CHAT})}: It provides the LLM with the PUT and the specification, then asks it to generate bug-identifying test cases directly.

\noindent $\bullet$ \textbf{Differential Prompting Plus (\texttt{DPP})}: As described in Section~\ref{sec:relatedWork}, Differential Prompting~\cite{nuance} is a state-of-the-art test case generation approach. While it is not designed or evaluated for detecting bugs in plausible programs, we recognize its potential in this area and use it as a baseline. The original method requires inferring the specification from the PUT; however, for a fair comparison, we use the ground truth specification instead. Thus, this modified version of Differential Prompting is referred to as Differential Prompting Plus (\texttt{DPP}).

\noindent $\bullet$ \textbf{Automated Program Repair (\texttt{APR}).} We introduce an additional baseline, which is an automated program repair (APR) method. This method provides the LLM with the PUT and specification, asking it to generate repair patches. In this scenario, correct patches are true positives, incorrect plausible patches are false positives, and other patches are negatives. This baseline corresponds to the first step of \ours, and we include it to show that \ours’s bug detection capability is not solely dependent on the LLM-generated repair patches.

\subsection{Implementation}
We use \texttt{gpt-3.5-turbo-0125} as the LLM for implementing \ours and the baselines, striking a balance between performance and cost to conduct our extensive evaluation within budget constraints.

\section{Results}

\begin{table*}[h]
\centering
\caption{(RQ3) Results of ablation study. ``PG'', ``IG'', and ``DT'' represent different ways to perform program variant generation, input generation, and differential testing, respectively. ``R'', ``P'', and ``F'' represent recall, precision, and F1 score, respectively.}
\label{tab:alation}
\begin{adjustbox}{width=\textwidth}
\begin{tabular}{llllrrr@{\hspace{3mm}}rrr@{\hspace{3mm}}rrr@{\hspace{3mm}}rrr@{\hspace{3mm}}rrr}
\toprule
\multirow{2}{*}{Pattern} &\multirow{2}{*}{ PG} & \multirow{2}{*}{IG} & \multirow{2}{*}{DT}& \multicolumn{3}{c}{k=2} & \multicolumn{3}{c}{k=4} & \multicolumn{3}{c}{k=6}& \multicolumn{3}{c}{k=8}& \multicolumn{3}{c}{k=10}\\ \cmidrule(l){5-7} \cmidrule(l){8-10}\cmidrule(l){11-13}\cmidrule(l){14-16}\cmidrule(l){17-19}
&& & & R & P & F & R & P & F& R & P & F& R & P & F& R & P & F\\ \midrule
 
1& Basic  &Basic & Basic  & .21 & .26 & .23 & .19 & .32 & .24 & .18 & .42 & .25 & .16 & .53 & .25 & .16 & .61 & .25
 \\ \midrule
2& Filtered  &Basic & Basic  & .20 & .27 & .23 & .19 & .34 & .24 & .17 & .45 & .25 & .16 & .50 & .25 & .16 & .48 & .24 \\ \midrule
3& Filtered  &Basic & Ours  &.23 & .60 & .33 & .22 & .59 & .32 & .22 & .59 & .32 & .22 & .59 & .32 & .22 & .59 & .32\\ \midrule
4&Ours  &Basic & Ours &.22 & .48 & .30 & .24 & .47 & .32 & .24 & .47 & .32 & .25 & .47 & .32 & .25 & .47 & .32
 \\ \midrule
5& Filtered &Ours & Ours & .26 & .76 & \textbf{.38} & .26 & .76 & .38 & .26 & .76 & .38 & .26 & .76 & .38 & .26 & .76 & .38 \\ \midrule
6& Ours  &Ours & Ours  & .25 & .70 & .37 & .28 & .70 & \textbf{.40} & .29 & .69 & \textbf{.41} & .29 & .69 & \textbf{.41} & .29 & .70 & \textbf{.41} \\ 
\bottomrule
\end{tabular}
\end{adjustbox}
\end{table*}

\subsection{RQ1: Performance on Buggy Plausible Programs}
\label{sec:res_find_defects}

Table~\ref{tab:effectiveness_all} shows the effectiveness of \ours and baselines in generating bug-identifying test cases for TrickyBugs and EvalPlus datasets. The last three rows highlight the improvements of \ours compared to the best baseline, \texttt{DPP}. To ensure a comprehensive comparison, we use three distinct methods: ``Average'', ``Best vs. Best'', and ``Worst vs. Worst.'' The ``Average'' comparison computes the mean value across all different $k$ values. ``Best vs. Best'' uses the $k$ values with the best F1 scores for comparison. For example, \texttt{DPP} achieves its best F1 score (24.95) on TrickyBugs (C++) at $k=8$, while \ours reaches its best F1 score (41.31) at $k=10$; we then compute the improvement of \ours ($k=10$) over \texttt{DPP} ($k=8$). ``Worst vs. Worst'' follows a similar approach but chooses the $k$ values with the worst F1 scores.

The evaluation results demonstrate \ours's superior performance in recall, precision, and F1 score, achieving up to $1.80\times$, $2.65\times$, and $1.66\times$ of \texttt{DPP}, respectively. Furthermore, \ours achieves the highest F1 score across all datasets. Specifically, \ours achieves F1 scores of 41.31\%, 42.35\%, and 51.34\% on TrickyBugs (C++), TrickyBugs (Python), and EvalPlus, respectively, significantly outperforming \texttt{DPP}'s F1 scores of 24.95\%, 36.20\%, and 35.76\%.

\finding{\ours shows the best effectiveness in detecting bugs in both human-written and AI-generated plausible programs. The recall, precision, and F1 score achieved by \ours are $1.80\times$, $2.65\times$, and $1.66\times$ those of the state-of-the-art baseline.}

\subsection{RQ2: Performance on Correct Programs}
\label{sec:res_on_correct_programs}
We further evaluate \ours and baseline methods on correct programs (i.e., canonical programs provided by the datasets) to assess whether they introduce false positives (FPs). We categorize the FPs into two types: incorrect oracles and invalid inputs.  Since \texttt{APR} does not generate test cases, it is excluded from this RQ. Given that EvalPlus provides official checkers for input validity, we focus on this dataset for RQ2.

\begin{figure}[h]
  \centering
  \includegraphics[width=1\linewidth]{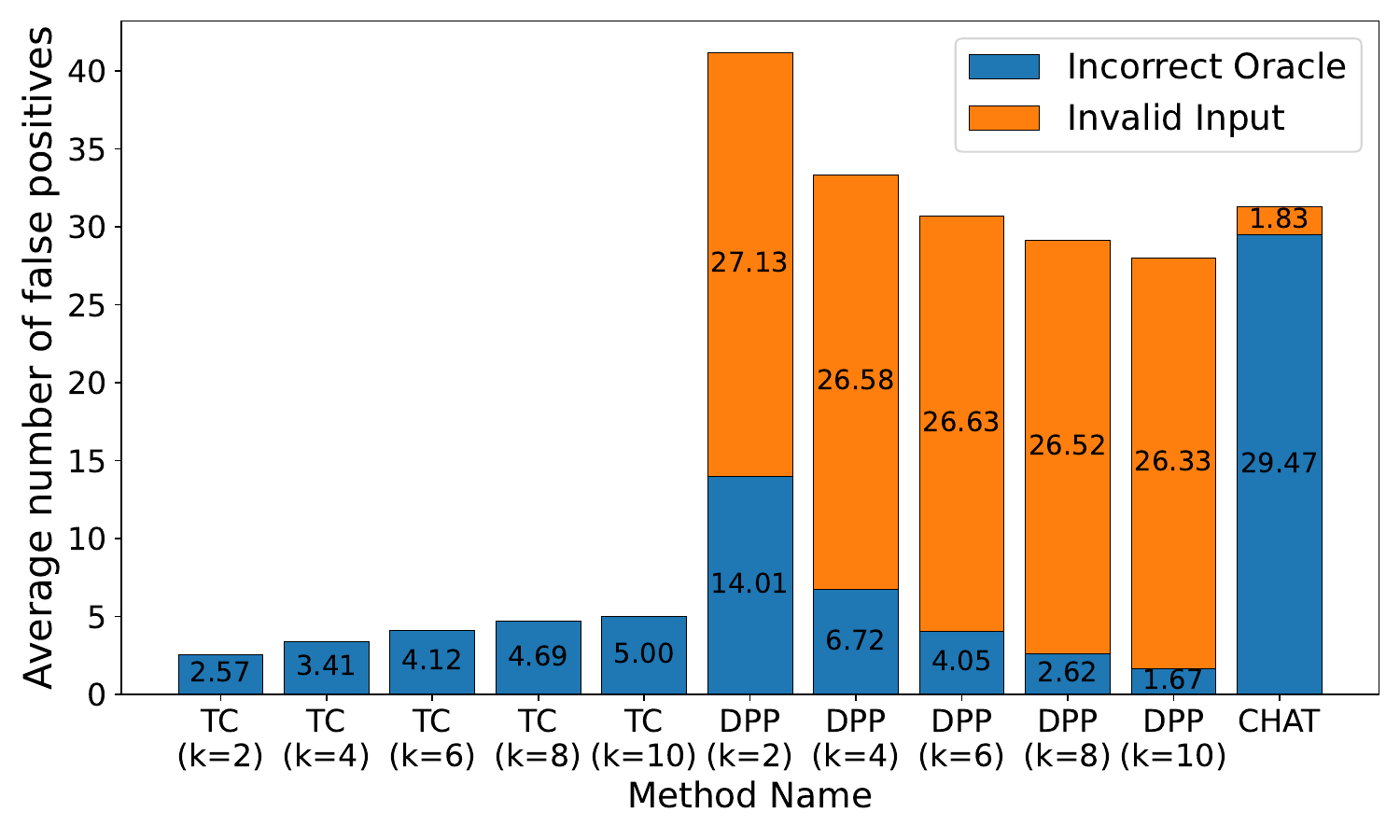}
  \caption{(RQ2) False positives generated by each approach for correct programs. Lower values indicate better performance. \ours generates significantly fewer false positives compared to the other methods.}
  \label{fig:can_res}
\end{figure}

Figure~\ref{fig:can_res} presents the results. The total number of FPs generated by \ours is significantly lower (up to 16$\times$) compared to \texttt{DPP} and \texttt{CHAT}. Notably, \ours produces no FP due to invalid inputs, demonstrating that its generator-based input generation method can effectively ensure valid inputs. In contrast, most FPs from \texttt{DPP} are due to invalid inputs. Additionally, the majority of FPs from \texttt{CHAT} are caused by incorrect oracles, highlighting that LLMs struggle to directly generate accurate test oracles for given test inputs.

\finding{\ours generates up to 16$\times$ fewer false positives for correct programs compared to state-of-the-art methods.}

\subsection{RQ3: Ablation Study}\label{sec:ablation}
\label{sec:res_ablation}

We conduct an ablation study to evaluate the contributions of different components to \ours's performance. Due to the page limit, we focus on TrickyBugs(C++) dataset.

Table~\ref{tab:alation} presents the results of the ablation study. In the table, ``PG'', ``IG'', and ``DT'' refer to different ways for program variant generation, input generation, and differential testing, respectively.

\noindent $\bullet$ For program generation, ``Basic'' generates program variants solely based on the specification and does not filter them using the existing test suite. The ``Filtered'' approach also generates variants in the same way as ``Basic'' but then filters them using the existing test suite. ``Ours'' refers to \ours's PUT-guided program generation.

\noindent $\bullet$ For input generation, the ``Basic'' approach directly uses the LLM to generates test inputs based on the specification, while ``Ours'' employs \ours's generator-based input generation method.

\noindent $\bullet$ For differential testing, ``Basic'' follows the majority voting rule for determining the test oracle. ``Ours'' implements \ours's diversity-driven differential testing.

Pattern 6 in the table represents the complete \ours approach.

The results demonstrate the great contribution of each component of \ours: PUT-guided program generation (by comparing Patterns 3, 5, and 6), generator-based generation (by comparing Patterns 4 and 6), and diversity-driven differential testing (by comparing Patterns 2 and 3).

\finding{Each component of \ours (i.e., PUT-guided program generation, generator-based test generation, and diversity-driven differential testing) contributes to its final performance.}

\begin{figure}[t]
    \centering
    \begin{subfigure}[b]{0.35\textwidth}
        \includegraphics[width=\textwidth]{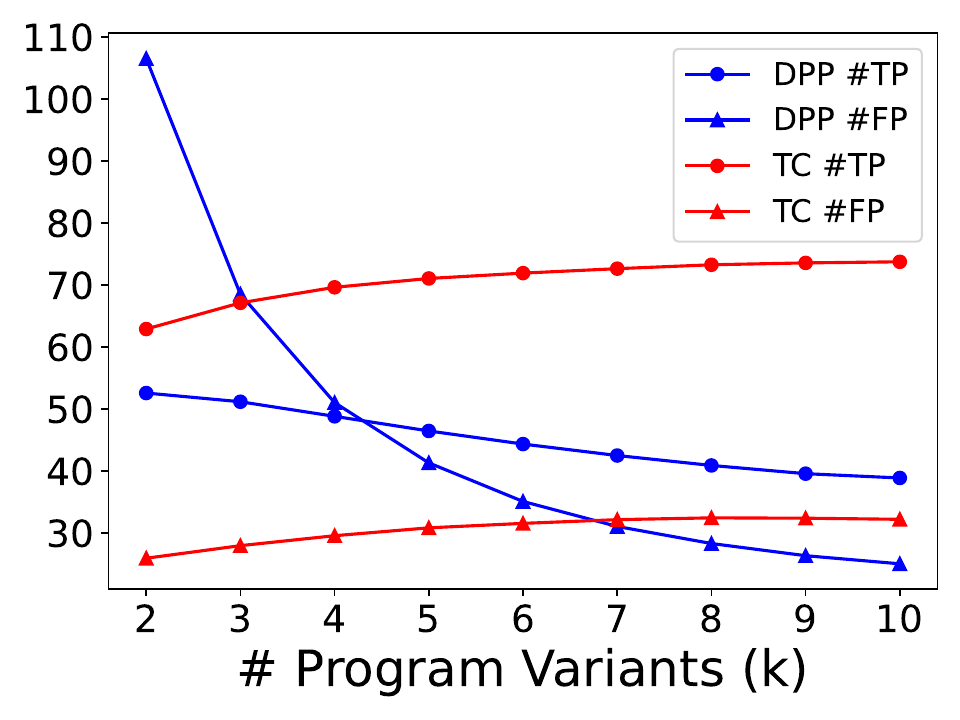}
        \label{fig:impact_of_k_TPFP}
    \end{subfigure}
    
    \begin{subfigure}[b]{0.35\textwidth}
        \includegraphics[width=\textwidth]{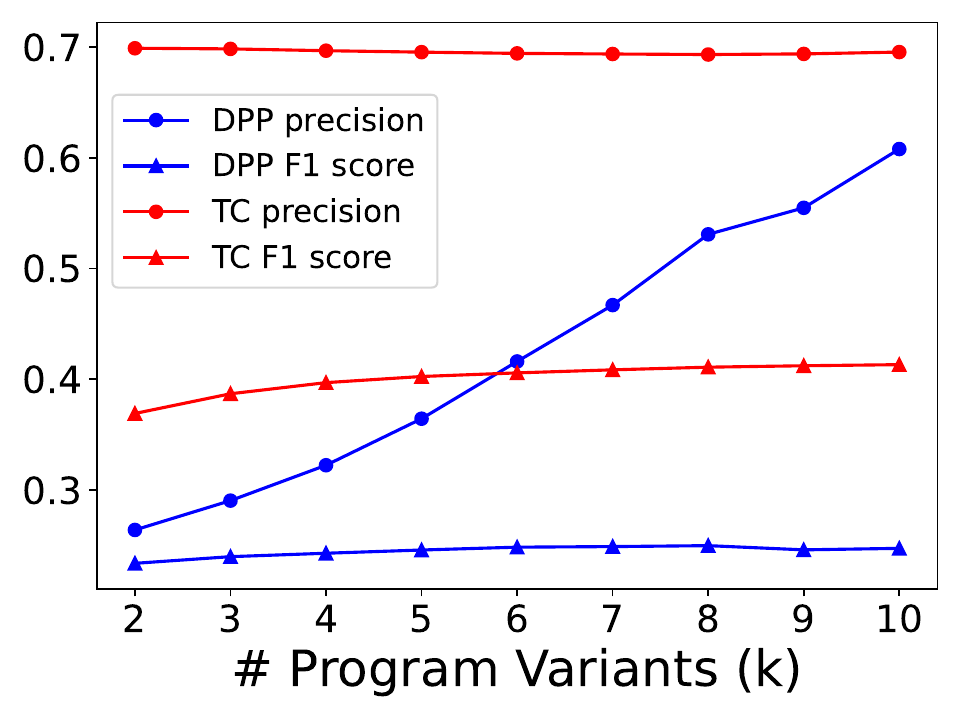}
        \label{fig:impact_of_k_preF1}
    \end{subfigure}
    \caption{(RQ4) Impact of the number of generated program variants $k$. \ours consistently maintains stable precision and F1 scores.}
    \label{fig:impact_of_k}
\end{figure}

\subsection{RQ4: Impact of Program Variant Number}
\label{sec:res_impact_of_k}
To better understand how the number of program variants $k$ impacts the effectiveness of \ours and \texttt{DPP}, We analyze the number of FPs, the number of TPs, precision, and F1 score as $k$ changes for the TrickyBugs (C++) dataset. 

Figure~\ref{fig:impact_of_k} shows the results. We can observe that \ours's TP, FP, precision, and F1 score are superior to \texttt{DPP} in most cases. For precision and F1 score, \ours, even in the worst case, outperforms \texttt{DPP} in the best case. Furthermore, the performance of \texttt{DPP} fluctuates significantly with changing $k$, while the performance of \ours remains consistently stable and excellent, further demonstrating the practicality of \ours.

\finding{The performance of \ours remains consistently stable and high with different numbers of program variants.}

\subsection{RQ5: Impact of Task Difficulty}
\label{sec:res_impact_of_difficulty}

\begin{figure}[t]
  \centering
  \includegraphics[width=1\linewidth]{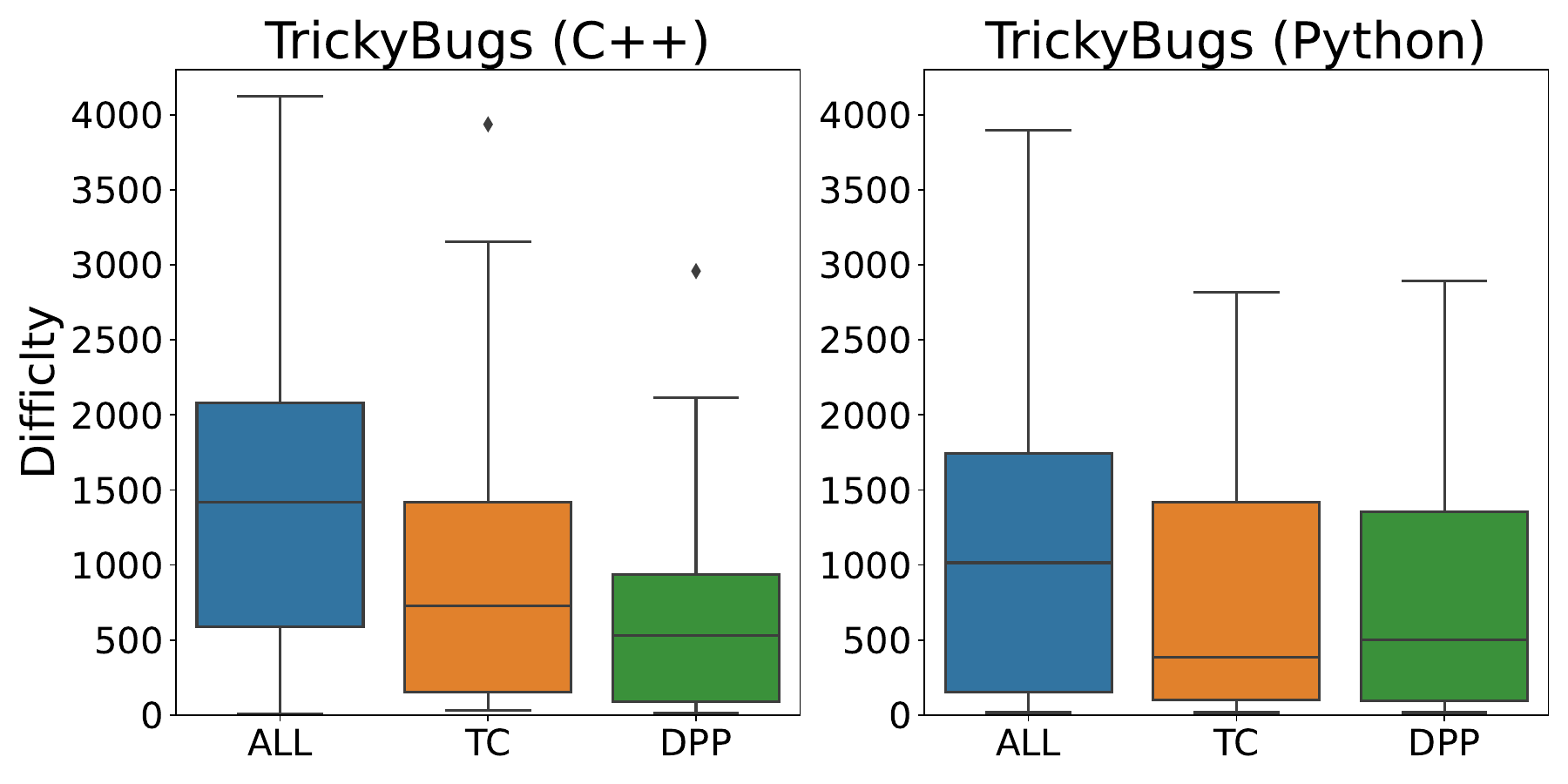}
  \caption{(RQ5) The distribution of the difficulty of all tasks and the tasks where \texttt{TC} and \texttt{DPP} perform well. }
  \label{fig:diff_compare}
\end{figure}

We further explore the impact of the difficulty of coding tasks on the effectiveness of \ours. Since the TrcikyBugs dataset provides difficulty information, we focus on it here.

\begin{figure}[t]
  \centering
  \includegraphics[width=1\linewidth]{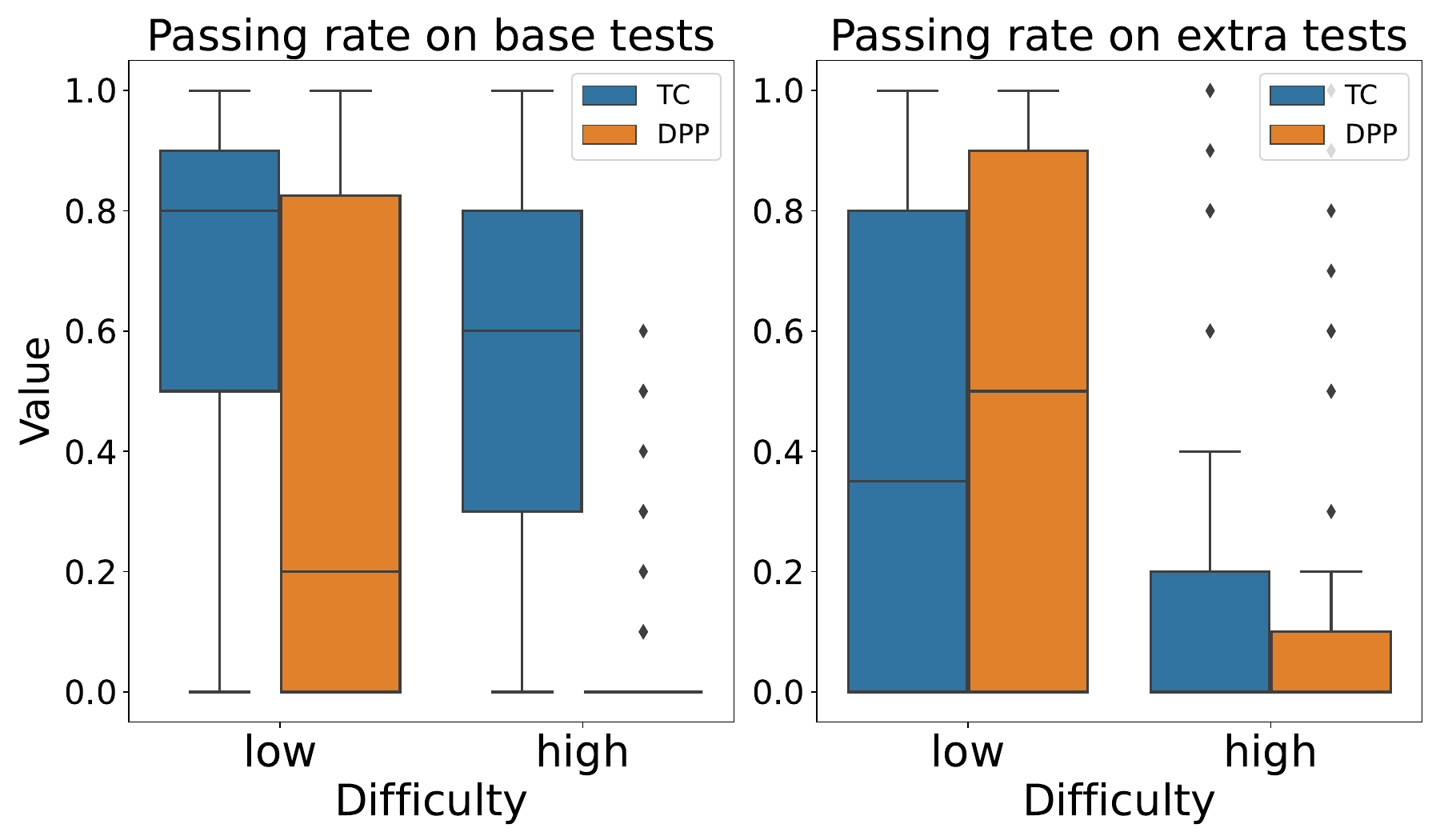}
  \caption{(RQ5) Average passing rate of the generated program variants on base test cases and extra test cases. The data is grouped by difficulty, ``low'' represents the lower 50\%, and ``high'' represents the higher 50\%.}
  \label{fig:pass_rate}
\end{figure}
 
Figure~\ref{fig:diff_compare} shows the results. ``ALL'' is the difficulty distribution of all tasks in the dataset. The next two box plots represent the difficulty distribution of the tasks where the corresponding method performs well (defined as achieving a precision greater than 0.5). 
We can find that when the difficulty of the coding task is low, the performance of the two methods is similar (see the right figure). However, for more difficult coding tasks, \ours outperforms \texttt{DPP} (see the left figure).
Figure~\ref{fig:pass_rate} shows another interesting result that more program variants generated by \ours pass the base tests, and this difference is more pronounced in high-difficulty tasks. This difference suggests that \ours introduces fewer new bugs when generating program variants, thereby improving the final performance.

\finding{\ours demonstrates a more significant improvement over \texttt{DPP} on the PUTs for more difficult coding tasks.}

\section{Discussion}

\subsection{Usefulness of Buggy Program Variants}

Here we discuss an interesting finding that \textbf{buggy program variants can also contribute to generating true positive test cases}. We refer to the variants that have produced the correct oracle for any true positive test case as \textit{useful variants}. During our evaluation, we find that 23.2\% (TrickyBugs) and 15.0\% (EvalPlus) of the useful variants are actually buggy (A program variant is buggy if it has ever produced any wrong output that is different from the canonical solution). 

The performance comparison between \texttt{TC} and \texttt{APR} in Table~\ref{tab:effectiveness_all} also supports the conclusion that buggy variants can be useful, as TrickCatcher achieves better recall than \texttt{APR} in most cases.

The findings imply that there is a certain logical complementarity between these buggy variants and the buggy PUT. So, even if a variant is not entirely correct, it may still contribute to generating a bug-identifying test case.
This observation aligns with the points we made in Section~\ref{sec:baseline} that the ability of \ours to detect bugs is not solely derived from LLM-based program repair, \ours can also leverage buggy variants to generate correct test oracles.

\subsection{Method Generalization Capability}

We also use another model, deepseek-v3, for experiments on EvalPlus dataset to verify the generalization capability of TrickCatcher. The evaluation results are shown in Table~\ref{tab:different_models}.

\begin{table}[h]
\centering
\caption{Evaluation results (EvalPlus) with different language models.}
\label{tab:different_models}
\begin{adjustbox}{width=\linewidth}
\begin{tabular}{lrrr} %
\toprule
Model & Recall & Precision & F1 score \\
\midrule
deepseek-v3(k=5)	 & 44.26 & 90.94 & 59.54\\
deepseek-v3(k=10)	 & 44.01 & 90.43& 59.21 \\
gpt3.5-turbo(k=5)	& 35.97 & 84.12 & 50.39 \\
gpt3.5-turbo(k=10) & 37.14 & 83.14 & 51.34 \\
\bottomrule
\end{tabular}
\end{adjustbox}

\end{table}

The experimental results demonstrate the generalization capability of TrickCatcher, and we can also find that the stronger the underlying model, the better the performance.

\section{Conclusion}
\label{sec:conclusion}
We propose \ours, an LLM-powered test case generation approach for detecting bugs in plausible programs. We evaluate \ours on both human-written and AI-generated plausible programs. The results show that \ours achieves up to 1.80×, 2.65×, and 1.66× the recall, precision, and F1 score of the state-of-the-art baseline, respectively. Additionally, the ablation study demonstrates that each component of \ours contributes to its performance.

\section*{Limitations}
The first limitation is that, due to budget constraints, we use two models, gpt-3.5-turbo and deepseek-v3, for evaluation. However, we believe that utilizing more advanced LLMs could further enhance the performance of \ours. The second limitation is the inherent uncertainty in the behavior of LLMs. To mitigate this, we performed multiple repetitions and averaged the results to ensure a more reliable evaluation. The third limitation concerns the risk of data leakage. However, the TrickyBugs dataset we used was released after gpt-3.5-turbo-0125, and EvalPlus explicitly prohibits its use for training LLMs. Moreover, the poor performance of the three LLM-based baselines further suggests that data leakage is not a main concern in our evaluation.

\section*{Acknowledgements}
This research is supported by the National Key R\&D Program under Grant No.2023YFB4503801, the National Natural Science Foundation of China under Grant No.62192733, 62192730, and the Major Program (JD) of Hubei Province (No.2023BAA024). Jie M. Zhang is supported by the ITEA Genius and ITEA GreenCode projects, funded by InnovateUK.

\bibliography{AID_refs}


\appendix

\section{Category of Test Cases}
\label{app:category_of_test_cases}

Based on the problem definition in Section~\ref{sec:prelimi}, we define a comprehensive category of test cases:

\textbf{\ding{182} Test cases that correctly identify a bug ($\boldsymbol{T_c}$).} For any $t = (in, out) \in T_c$, we have \( f(in) = out \) and \( PUT(in) \neq out \). This is exactly the test case we want, which effectively identifies a bug in PUT.

\textbf{\ding{183} Test cases with right test oracles but do not identify any bug ($\boldsymbol{T_r}$).} For any \( t = (in, out) \in T_r \), we have \( f(in) = out \) and \( PUT(in) = out \). Test cases of this kind are trivial, neither positively nor negatively significant for bug identification.

\textbf{\ding{184} Test cases with wrong test oracles ($\boldsymbol{T_w}$).} For any \( t = (in, out) \in T_w \), we have \( f(in) \neq out \). These test cases are erroneous. They may lead to \textit{false negative}, where \( PUT(in) = out \) but \( PUT(in) \neq f(in) \). They may also lead to \textit{false positive}, where \( PUT(in) \neq out \) but \( PUT(in) = f(in) \). False positives are more detrimental than false negatives because false positives can undermine the credibility and practicality of the entire method.

\textbf{\ding{185} Test cases with invalid input ($\boldsymbol{T_{err}}$).} For any \( t = (in, out) \in T_{err} \), we have \( in \notin \text{input space } I \). These test cases are erroneous. For any invalid test input, no test oracle exists, and all program behaviors are undefined. These test cases might also lead to false positives.

Any test case should fall into one of the above four categories.

For convenience, we define two additional types of test cases: We define \textbf{passed test cases ($\boldsymbol{T_p}$).} For any \( t = (in, out) \in T_p \), we have \( PUT(in) = out \). A passed test case $t$ could belong to $T_r$, $T_w$, or $T_{err}$. Then we define \textbf{failed test cases ($\boldsymbol{T_f}$).} For any \( t = (in, out) \in T_f \), we have \( PUT(in) \neq out \). A failed test case $t$ could belong to $T_c$, $T_w$, or $T_{err}$.

We focus on identifying functional bugs within plausible programs; therefore, we do not consider other program behaviors such as timeouts or crashes, which fall outside the scope of our study.

\section{Experiment Repetition}
\label{sec:repetition}
To make our evaluation results more reliable, we repeat the experiments multiple times to reduce the impact of randomness from LLMs.

For \texttt{CHAT}, each response of the LLM is a test case. We repeatedly sample 100 test cases and compute the average number of TPs and FPs.

For \texttt{APR}, each response of the LLM is a patch. We repeatedly sample 10 patches and compute an average number of TPs and FPs.

For \texttt{DPP} and \texttt{TC}, the randomness comes from program generation and input generation. We first repeatedly sample 100 inputs and 10 program variants, then keep only the filtered program variants (if the method in the first step is \texttt{TC} or filtered). Next, we use a combinatorial approach to conduct extensive repeated experiments. For example, we have 10 filtered program variants for a PUT and set the parameter $k$ (number of generated programs) as 4. Furthermore, we randomly select 4 out of the 10 program variants each round, resulting in a total of $C_{10}^{4}=210$ rounds of results. For each round, we first compute the average number of TPs and FPs among the 100 generated inputs, getting an average precision and recall for this round. We average the results of the 210 rounds to obtain the final result.

\end{document}